\documentclass{emulateapj}

\font\tenbg=cmmib10 at 10pt
\def \rvecmu{{\hbox{\tenbg\char'026}}}
\def \rvecphi{{\hbox{\tenbg\char'036}}}
\def \rvectheta{{\hbox{\tenbg\char'022}}}

\font\tenbg=cmmib10 at 10pt

\def \rvecphi{{\hbox{\tenbg\char'036}}}

\begin{document}
\title{Screening of the Magnetic Field of
Disk Accreting Stars}

\author{R.V.E. Lovelace, M. M. Romanova, and}
\affil{Department of
Astronomy, Cornell University, Ithaca, NY
14853; RVL1@cornell.edu;
romanova@astro.cornell.edu}
\author{G.S. Bisnovatyi-Kogan}
\affil{Space Research Institute,
Russian Academy of Sciences,
Moscow, Russia; gkogan@mx.iki.rssi.ru}

\begin{abstract}

     An analytical
model is developed for  the screening of
the external magnetic field of
a rotating, axisymmetric neutron star
due to the  accretion of plasma
from a disk.
     The decrease of the
field occurs   due to the electric current
in the infalling plasma.
   The deposition of this
current carrying plasma
on the star's surface  creates
an induced magnetic moment
with a sign opposite to that of the
original magnetic dipole.
   The  field decreases
independent of whether the star
spins-up or spins-down.
    The time-scale for an
appreciable decrease (factor of
$>100$) of the field is
found to be $\sim 1.6 \times 10^7$ yr,
for a mass accretion rate $\dot{M}=10^{-9} M_\odot/$yr
and an initial magnetic moment
$\mu_i = 10^{30}{\rm~\! G cm}^3$ which
corresponds to a surface field of $10^{12}~\!$G if
the star's radius is $10^6$ cm.
    The time-scale varies approximately as
$\mu_i^{3.8}/\dot{M}^{1.9}$.
   The decrease of the magnetic field
does not have a simple relation to
the accreted mass.
   Once the accretion stops the field
leaks out on an Ohmic diffusion time scale
which is estimated to be $ > 10^9$ yr.

\end{abstract}

\keywords{stars: neutron --- pulsars: general---
---  stars: magnetic fields --- X-rays: stars}

\section{Introduction}

    The decrease of the external magnetic field
of accreting neutron stars has been
a long standing puzzle and has been
explained as being due to Ohmic
decay of the field
(e.g., Goldreich \& Reisenegger 1992),
crustal motion
on the star's surface  (Ruderman 1991), or
``burial'' or screening of the original
magnetic field by the accreted matter
(Bisnovatyi-Kogan \& Komberg
1974).
    This work develops an analytic
model for the decrease of the external
field due to accretion of plasma.

   For some time after the discovery of
pulsars only single stars
objects were found.
    It appeared
that pulsars did not occur in binaries.
    Because more that half of
all stars are in binaries, the
occurrence of  isolated pulsars
was explained either by pair
disruption during supernova
explosion leading to pulsar formation,
or by the absence of SN
explosions at the end of evolution of
stars in close binaries
(Trimble \& Rees 1971).
     Bisnovatyi-Kogan and Komberg
(1974; hereafter BK74)
analyzed the evolution of
X-ray binaries in low-mass
systems (e.g., Her X-1) and
    concluded that evolution of
such systems should lead
to the formation of
non-accreting neutron star, i.e., radio
pulsars, in close binary systems.
    BK74 showed
that the neutron star rotation is
accelerated during the disk accretion
stage so that a reborn
(or recycled) radio pulsar should
become  detectable provided its
magnetic field is similar to that of
isolated pulsars.
    The absence of radio pulsars
in close binaries in extensive
searches could be explained
by  only one reason (BK74):
During the accretion stage the magnetic
field of the neutron star is
screened by the inflowing plasma,
so that the recycled pulsar
should have $B\sim 10^8-10^{10}$ G,
which is $2-4$ orders of magnitude
smaller than the field strength of
radio pulsars.
    Discovery of
the first binary pulsar by Hulse and
Taylor (1975), and subsequent
discovery of more than 50 recycled
pulsars (Bhattacharya \& van
den Heuvel 1991; Lorimer 2001; Lyne
et al. 2004)  confirmed
the conclusion that  recycled pulsars
have small magnetic fields as  predicted by BK74.

\begin{figure*}[t]
\epsscale{1.0}
\plotone{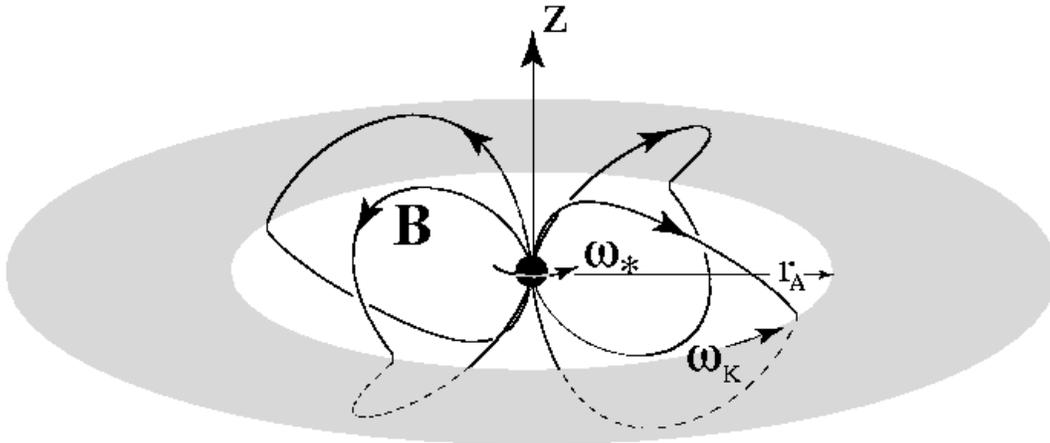}
\caption{Three-dimensional view of four
field lines going from the star to the
Alfv\'en radius $r_A$.
    For the conditions shown
the disk angular velocity at
$r_A$, $\omega_K(r_A)$, is larger
than the star's angular velocity $\omega_*$.
}
\end{figure*}

   From a statistical analysis
of $24$ binary radio pulsars with nearly
circular orbits and low
mass companions,
Van den Heuvel and Bitzaraki (1995)
discovered a clear
correlation between the spin
period $P$ and the orbital period
$P_{orb}$, as well as between the
magnetic field and the orbital period.
The pulsar period and magnetic
field strength increase with the
orbital period at $P_{orb}>100$
days, and scatters around $P\sim 3$ ms
and $B\sim 2\times 10^8$
G for smaller binary periods.
    These relations strongly suggest
that an increase in the amount of
accreted mass leads to a screening
of the initial magnetic field,
and that there is a lowest
field strength of about
10$^8$ G.
   Magnetic field
screening during accretion has been
discussed by a number of authors
(see, e.g., BK74; Romani 1990; Wijers 1997;  Cheng \&
Zhang 1998, 2000;  Choudhuri \& Konar 2002;
Payne \& Melatos 2004).
   The studies by Cheng and Zhang and by
Payne and Melatos analyze the
flow and magnetic field
evolution {\it within} the neutron
star in contrast with the present work which
is concerned mainly with the magnetosphere
of the star.

   Here we present an analytical
model of the screening of
the magnetic field of
a rotating neutron star
due to the accretion of plasma
from a disk.  The system is
assumed to be axisymmetric.
    In the idealized
case of a non-conducting sphere, the
decrease of the external magnetic
field occurs   due to the electric current
in the infalling plasma.
   The accretion of this
current carrying plasma
to the star's surface  creates
an induced magnetic moment
with a sign opposite to that of the
original magnetic dipole.
   In the more
realistic case where the
star is treated as a conducting
sphere, the magnetic field due
to the accreted current-carrying plasma
does not penetrate the main volume
of the star.
     As a result the screening effect
of the  plasma is
greatly reduced (compared with
the case of a non-conducting star), and the time-scale
for appreciable screening is greatly
increased.
     For representative parameters
we find that the magnetic field
of the conducting star decreases
significantly on a time-scale of
the order of $10^7$ yr.
    The screening mechanism stops working
when the Alfv\'en radius is comparable to
the radius of the star.

In the \S 2 the general picture is
described, and the main equations are
derived. In \S 3 and \S 4 the magnetic
field decrease during plasma accretion
is considered for non-conducting and
conducting spheres, respectively.
We calculate the time-scale for an
appreciable decrease of the magnetic
field of a conducting star which
depends on the radial thickness of
the accreted matter.
   In \S 5 we
estimate the thickness of the
accreted matter and the time scale
for the magnetic field to diffuse out
of the star once accretion has ceased.
    In \S 6 we give the conclusions of
this work.

\section{Theory}

    We consider disk accretion to a
rotating neutron star with
an aligned dipole magnetic field as indicated
in Figure 1.
      That is, we consider an {\it axisymmetric} star,
magnetic field, and disk.
     Further, we consider configurations which
are mirror symmetric about the equatorial plane.
     We use both spherical $(R,\theta, \phi)$
and cylindrical $(r,\phi,z)$ non-rotating
coordinate systems.

       For this system the
{\it corotation radius} is
\begin{equation}
r_{cr} = \left({G M \over \omega_*^2}\right)^{1/3}
\approx 1.7\times 10^8~ P^{2/3}~{\rm cm}~,
\end{equation}
where, $\omega_*$ is the angular rotation rate
of the star, $P=2\pi/\omega_*$  is its period,
and the star's mass is considered to be $1.4M_\odot$.
The {\it Alfv\'en  radius} $r_A$ (or inner radius)
of the disk is the distance at which
the kinetic energy density of
the matter $\rho {\bf v}^2/2$
is comparable with the magnetic energy
density ${\bf B}^2/8\pi$ (Ghosh \& Lamb 1979).
This gives
\begin{eqnarray}
r_A &\approx& k_A\left( {\mu^2
\over  \dot{M}\sqrt{GM}}\right)^{2/7}~,
\nonumber \\
&\approx& 1.94 \times 10^8 \left({k_A \over 0.5}\right)
\left({\mu_{30}^2
\over \dot{M}_{-9}}\right)^{2/7}~{\rm cm}~.
\end{eqnarray}
Here, $\mu$ or $\mu_{30}\equiv
\mu/(10^{30}$ G~cm$^3$) is the star's magnetic
moment,
$\dot{M}$ or $\dot{M}_{-9} \equiv
\dot{M}/(10^{-9} M_\odot/{\rm yr})$ is the mass
accretion rate.
    The accretion luminosity
is $L_{accr}=GM\dot{M}/R_* \approx
1.2\times10^{37}\dot{M}_{-9}$ erg/s
assuming the star's radius $R_*=10^6$ cm,
while the Eddington luminosity for
a $1.4M_\odot$ star is $L_{Edd}\approx
1.76\times 10^{38}$ erg/s.
    Thus  $L_{accr}/L_{Edd}=\approx 0.067\dot{M}_{-9}$.
      The dimensionless
coefficient $k_A$ is of order of one-half and
depends weakly on the disk parameters
(see Ghosh \& Lamb 1979;
  Lovelace, Romanova, \& Bisnovatyi-Kogan
1995;  Long, Romanova, \& Lovelace 2004).
The star's accretion
luminosity is $\dot{E}_a=G M \dot{M}/R_*
\approx 1.2\times 10^{37} \dot{M}_{-9}$ erg/s  with
the star's radius $R_*$ assumed to be $10^6$ cm.

      The ratio $r_A/r_{cr}$ determines the
qualitative evolution of the system.
    For $r_A \lesssim r_{cr}$, accretion
causes  the star to {\it spin-up}
   with  the  rate
of increase  angular momentum
$d J/dt \approx \dot{M}(GM r_A)^{1/2}$,
where $J=I\omega_*$, with $I=10^{45} I_{45}$ g cm$^2$
the star's moment of inertia.
     On the other hand for $r_A \gtrsim r_{cr}$,
the star {\it spins-down} with the rate
  $dJ/dt \approx -\dot{M}(GM r_A)^{1/2}$.
   For $r_{cr} \ll r_A$ the star spins down
but  is in the
{\it propeller regime} where an appreciable fraction
of the accreting matter  may be
expelled from the system
  by the rotating magnetic field
(Illarionov \& Sunyaev 1975;  Lovelace, Romanova,
\& Bisnovatyi-Kogan 1999) or
the accretion to the star's surface
may be highly non-stationary  (Romanova
et al. 2004).  In the propeller regime,
the mass accretion rate in the outer disk
($\dot{M}$ in equation 2) may be substantially
larger than the mass accretion rate to the
surface  of the star  (Lovelace, et al. 1999).
   We consider only cases where $r_A > R_*$,
with $R_*$ the star's radius.

     For distances $R$ less than $r_A$, the
accreting matter moves in a {\it funnel
flow} which follows approximately
the dipole magnetic lines as sketched in Figure 2 (see, e.g.,
Romanova et al. 2002).
     For $R \ll r_A$ the velocity of the
matter in the funnel flow is approximately
free-fall.
     The cross sectional area
of the funnel flow $S$ varies as $R^3$
for a dipole field so that
$\rho \propto 1/R^{5/2}$.
Consequently,
$\rho {\bf v}^2/2$ varies as $1/R^{7/2}$.
     Thus the kinetic energy density of the
funnel matter is much less that that of the
magnetic field which varies as $1/R^6$.
     Therefore, the funnel flow
is magnetically dominated
or {\it force-free} for $R < r_A$.

\begin{figure*}[t]
\epsscale{0.8}
\plotone{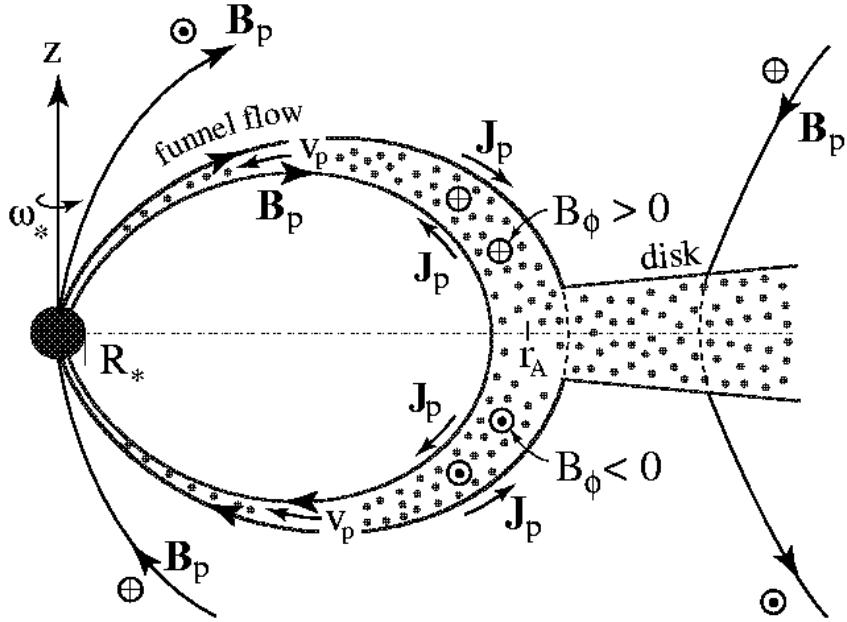}
\caption{Diagram of the poloidal cross-section
of the accretion disk and funnel flow.
${\bf B}_p$ is the poloidal magnetic field, $B_\phi$
is the toroidal magnetic field,
${\bf J}_p$ the poloidal current density,  and ${\bf v}_p$
is the poloidal component of the
funnel flow velocity.  The symbol $\odot$ represents
an arrow out of the page, while $\oplus$ represents
the opposite direction.  The solid dots indicate
a high plasma density.
}
\end{figure*}

     For the considered axisymmetric system,
the magnetic field has
the form
$$
   {\bf B}~ = {\bf B}_p +
B_\phi \hat{\rvecphi~}~,
$$
with
${\bf B}_p = B_R{\hat{\bf R}}+
B_\theta \hat{\rvectheta~}.$
We can write
$$
B_R =
{1 \over R^2\sin\theta}{\partial \Psi \over
   \partial \theta}~,\quad
B_\theta =-~{1 \over R\sin\theta}
{\partial \Psi \over \partial R}~,
$$
where $\Psi(R,\theta)
\equiv R\sin\theta A_\phi(R,\theta)$ is referred
to as the `flux function.'   Note that
$\Psi(R,\theta)=$ const is the equation
for the  poloidal projection of a magnetic
field line.

     In the
force-free limit,
   the flow speeds are
sub-Alfv\'enic,
${\bf v}^2 \ll {v}_A^2 = {\bf B}^2/4\pi \rho$,
where $v_A$ is the Alfv\'en velocity.
In this limit,
$
   {\bf J \times B} \approx 0$;
     therefore, ${\bf J} = \lambda {\bf B}$
(Gold \& Hoyle 1960).
    Because ${\bf \nabla \cdot J}=0$,
$({\bf B \cdot \nabla}) \lambda =0$
and consequently $\lambda = \lambda(\Psi)$.
     Thus, Amp\`ere's equation becomes
$
{\bf \nabla \times B} =
{4 \pi \lambda(\Psi)} {\bf B}/c$.
     The $R$ and $\theta$ components of Amp\`ere's
equation  imply
$$
   H(\Psi)= R\sin\theta B_\phi ~,\quad
{\rm and}\quad {dH \over d\Psi}
={4\pi \over c}
\lambda(\Psi)~,
$$
where $H(\Psi)$ is another function of
$\Psi$.
     Thus, $H(\Psi)=$ const along any
given  field line, and
${\bf J}_p=(c / 4\pi)
(dH / d\Psi){\bf B}_p$ so that
$({\bf J}_p \cdot{\bf \nabla}) H=0$.
     The toroidal component of Amp\`ere's
equation gives
\begin{equation}
\Delta^\star \Psi = -
H(\Psi) {d H(\Psi) \over d\Psi}~,
\end{equation}
where
$$
   \Delta^\star \equiv
{\partial^2 \over \partial R^2}
+{1-\zeta^2\over R^2}{\partial^2 \over \partial \zeta^2}~,
$$
and $\zeta \equiv \cos \theta$.
This is the Grad-Shafranov equation for
$\Psi$ (see e.g. Lovelace {\it et al.} 1986).
     Note that the lines $\Psi(R,\theta)=$const
are the poloidal magnetic field lines.
Alternatively the rotation of such a line
around the $z-$axis form a {\it flux surface}.

     The magnetic moment of the star,
$$
\rvecmu = {1 \over 2c} \int_{R\leq R_*}
d^3x ~{\bf r}\times {\bf J} = \mu~\hat{\bf z}
$$
where
\begin{equation}
\mu={\pi \over c}\int_0^{R_*} R^3dR ~
\int_0^\pi d\theta\sin^2\theta
~J_\phi~.
\end{equation}
    Initially, before there
has been appreciable accretion, we
assume that the magnetic moment $\mu_i=\mu(t=0)$
is due to current flow deep inside the star.
     Thus the star's initial magnetic field is
given by $\Psi_i(R,\theta) = \mu_i \sin^2\theta/R$
for $R \geq R_*$.   Note that $\Delta^* \Psi_i =0$.

      In the following we
show that accretion of current carrying
matter to the surface of the star acts to reduce
the star's magnetic moment.
      We consider first the case of
a non-conducting star in \S 3.
In \S 4 we consider the realistic
   case of a highly conducting star.

\begin{figure*}[t]
\epsscale{1.1}
\plotone{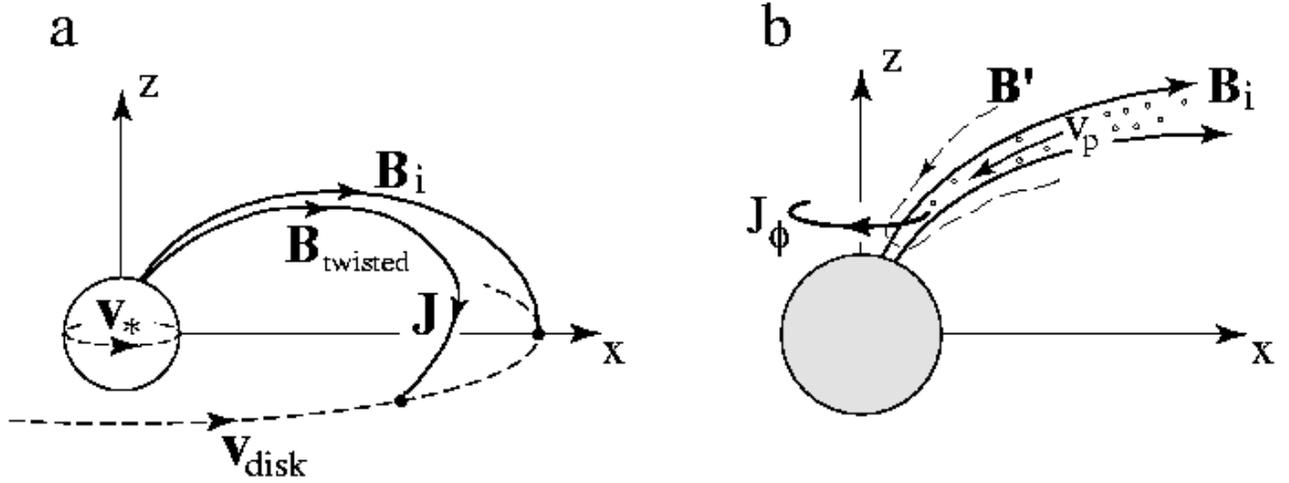}
\caption{({\bf a}) Schematic drawing of an initial
dipole  field line ${\bf B}_i$ (in
the $x,z$ plane) and the corresponding
twisted field line for the case where the
angular velocity of the disk at the Alfv\'en
radius $\omega_K(r_A)$ is {\it less} than the angular
velocity of the star $\omega_*$.
    This is opposite to the case shown in
Figures 1 and 2.
     The current flow ${\bf J}$ parallel
to the ${\bf B}$-field necessary
to produce this twist is into the star
at high latitudes and out of the star
at low latitudes.
     The low-latitude current flow
has the dominant  screening
effect and only this current is shown
in (${\bf a}$).
    This current flow along the
twisted field line gives rise to a
{\it negative} toroidal current density
$J_\phi$ near the star as shown in ($({\bf b}$).
  Panel ({\bf b}) shows the poloidal field
component  ${\bf B}^\prime$ arising from the
toroidal current $J_\phi$.
   The funnel flow is indicated by the small
circles and the arrow  ${\bf v}$ indicates the
flow velocity.
}
\end{figure*}

\section{Accretion to a Non-Conducting Star}

    This subsection  treats
the case of a non-conducting
star in order to explain our
model of the magnetosphere.
    This model star has a non-varying
dipole field source at its center.
   The magnetic field associated with
the accreting matter is allowed
to penetrate the star's surface $R=R_*$.

    The accreting plasma outside the
star is highly conducting.
      Consequently,
plasma falling onto the star
carries with it an embedded magnetic
field and the associated current density
as sketched in Figures 2 and 3.
     This plasma infall
causes the star's magnetic moment  to
change at the rate
\begin{equation}
{d \mu \over dt} ={\pi \over c} R_*^3 \int_0^\pi
d\theta \sin^2\theta ~\big( J_\phi v_R\big)_{R_*+0}~.
\end{equation}
    Here,  $R_*+0$ indicates a distance just outside
of the star, and
   $v_R$ is the plasma infall speed
which is approximately the free-fall speed
$v_{R*}=(2GM/R_*)^{1/2}$, because the
star's radius $R_*$ is considered
to be significantly
smaller than the
Alfv\'en radius $r_A$.
     This speed is much larger than the
rotational velocity of the star's surface
$\sim \omega R_*$, because the corotation
radius is also considered to be significantly
larger than $R_*$.
    We assume that there is a strong shock just
outside the star's surface which thermalizes
the bulk kinetic energy of the flow.

      The total magnetic field $\Psi
=\Psi_s+\Psi_m$ is
the sum of the field due to
the star, $\Psi_s$, and that
due to current flow in the magnetosphere,
$\Psi_m$.
      For the considered conditions where $R_* \ll r_A$,
we show below that $|\Psi_m| \ll \Psi_s$.
    We have $\Psi_s = \mu \sin^2\theta/R$ far from
the star,
which satisfies the
   $\Delta^* \Psi_s =0$.
     Consequently, equation (3)
gives $\Delta^* \Psi_m = -H(\Psi_s)
dH(\Psi_s)/d\Psi_s$.
      Because $\Psi_m$ is small,
   $\Psi$ (without the $s$
subscript) in subsequent analysis refers
to the field of the star.

     The funnel flow as shown in Figure 2
exists close to the flux surface $\Psi = \Psi_{ff}
= \mu /r_A$.
    For a dipole field the funnel flow hits
the star's surface at $\theta_{ff} \approx (R_*/r_A)^{1/2}$.
     The width in $\Psi$ of the funnel flow is
$\Delta \Psi_{ff}/\Psi_{ff}
\approx \Delta r_A/r_A \ll 1$.
    The radial width is assumed to
be of the order of the
half-thickness of the disk $h_A$
at the Alfv\'en radius $r_A$ as
indicated by MHD simulations of
funnel flows by Romanova et al. (2002).
        It is clear that $H(\Psi)$ is zero
outside of the funnel flow and that $H^2$ is
a maximum $H_m^2$ in  the middle of the funnel flow
at $\Psi=\Psi_{ff}$.
     The value of $H_m^2$ is
estimated by the condition that $B_\phi^2
\approx B_p^2$ at the Alfv\'en radius $r_A$.
For larger values of $(B_\phi/B_p)^2$ at $r_A$,
the magnetic
field loop would open (Lynden-Bell \& Boily 1994;
Lovelace, Romanova, \& Bisnovatyi-Kogan 1995).
     Thus $H_m^2 \approx \mu^2/r_A^4$.
This value can be used to show
that $|\Psi_m|/\Psi_s \approx (R_*/r_A)^4 \ll1$.

    In equation (5) we can use the fact that
\begin{equation}
J_\phi = { c \over 4 \pi R \sin\theta}~
{d  \over d \Psi} {H^2 \over 2}~.
\end{equation}
Thus
\begin{equation}
{d \mu \over dt} ={1 \over 4}
R_*^2~ v_{R*} \int_0^{\pi/2}
d\theta ~\sin\theta ~
{d  H^2\over d \Psi}\bigg|_{R_*+0} ~,
\end{equation}
where we have  combined the contributions
of the two hemispheres.
    On the star's surface we have
$\Psi=\mu \sin^2\theta/R_*$
so that $\cos\theta =(1-R_*\Psi/\mu)^{1/2}$.
    Thus we obtain
\begin{equation}
{d\mu \over dt}
={R_*^3 v_{R*} \over 8 \mu} \int_0^{\mu/R_*}
{d {\Psi} \over (1-R_*{\Psi}/\mu)^{1/2}}
~{d H^2 \over d {\Psi}}~.
\end{equation}
The square root can be  Taylor
expanded in that
$R_*{\Psi}/\mu \approx R_*/r_A$.
An integration by parts then gives
\begin{equation}
{d\mu \over dt} = -~{R_*^4 v_{R*} \over 16 \mu^2}
\int_0^{\mu/R_*} d {\Psi} ~H^2~.
\end{equation}
    Thus the star's magnetic moment decreases
independent of the sign of $H$ and
hence independent of the sign of
$\omega_K(r_A) -\omega_*$.

   Using the mentioned estimates
$H_m^2 \approx \mu^2/r_A^4$
and $\Delta \Psi_{ff} \approx (h_A/r_A)\Psi_{ff}
=(h_A/r_A)(\mu/r_A)$, equation (9)
gives
\begin{equation}
{d\mu \over dt} = -~{1\over 16}~\mu
\left({R_* \over r_A}\right)^4
\left({h_A \over r_A}\right)
\left( {v_{R*} \over r_A}\right)~.
\end{equation}
     This equation is not
realistic for a neutron star which
is highly conducting.

\section{Accretion to Conducting Star}

    We now treat the case of a highly
conducting star.
    For $R \leq R_*$ the
star is considered to be perfectly
conducting with the frozen in dipole
magnetic field $\Psi_i = \mu_i\sin^2\theta/R$.
   The magnetic field associated with the
accreting matter {\it cannot} penetrate
the surface $R=R_*$.
    The  shell from $R=R_*$ to $R_{a}>R_*$ is
the  layer of accreted matter.
     The layer thickness is
$R_{a}-R_* =\Delta R_a \ll R_*$.
    The accumulated azimuthal current
carried by the accretion layer
is modeled by surface current
layer ${\cal K}_\phi(\theta)$ at
$R_{\cal K}=R_*+ \Delta R_a/2$.

     The magnetic field in the different
radial regions can be expanded in
terms of solutions of $\Delta^* \Psi=0$,
because the contribution of the
magnetospheric currents is negligible.
    For the region $R > R_{\cal K}$
we have
\begin{equation}
\Psi^+ = \Psi_i(R,\zeta)+ \sum_{n=1,3,..}
a_n { G_{n}(\zeta) \over R^n}~,
\end{equation}
where $G_n(\zeta)$ is a Gegenbauer polynomial,
$\zeta \equiv \cos\theta$, and the $a_n$ are
coefficients determined below.
      The terms with even values of $n$ are
excluded because they correspond to
a field which is not mirror symmetric
about the $z=0$ plane.
    We can take
\begin{equation}
G_n(\zeta) =
(1-\zeta^2){d P_n(\zeta) \over d \zeta}~,
\end{equation}
where $P_n(\zeta)$ is the usual Legendre
polynomial.
    The other  solutions
related to the Legendre $Q_n$ functions
are unphysical.
      One can readily show that
the $G_n$ obey the
orthogonality relations
\begin{equation}
\int_{-1}^1 d \zeta ~{G_n(\zeta)G_m(\zeta)
\over 1-\zeta^2} = {2n(n+1)
\over 2n+1}~\delta_{nm}~.
\end{equation}
    For example,
$$
G_1=1-\zeta^2~,
$$
$$
G_3=-{3\over 2}+9\zeta^2-{15\over 2}\zeta^4~,
$$
and
$$
G_5={15\over8}-{225\over 8}\zeta^2+{525 \over 8}\zeta^4
-{315\over 8}\zeta^6~.
$$
     As $\zeta^2 \rightarrow 1$,
$G_n \rightarrow n(n+1)(1-\zeta^2)/2$.
     As $\zeta \rightarrow 0$ (the equatorial
plane), $G_n \rightarrow \sin(\pi n/2)n (n-2)!!/(n-1)!!$,
with values of $1, -3/2,$ and $15/8$ for $n=1,3,$ and $5$.

For $R_* \leq R < R_{\cal K}$, we have
\begin{equation}
\Psi^- = \Psi_i(R,\zeta)+ \sum_{n=1,3,..}
b_n \left({1\over R^n}-{R^{n+1} \over R_*^{2n+1}}
\right)
G_{n}(\zeta) ~,
\end{equation}
so that $ \Psi(R_*,\zeta) = \Psi_i(R,\zeta)$.
For $R\leq R_*$, $\Psi=\Psi_i(R,\zeta)$.

     Matching the inner and outer flux functions
at $R=R_{\cal K}$ gives
\begin{equation}
a_n=b_n\left[1-
\left({R_{\cal K} \over R_*}
\right)^{2n+1}\right]~.
\end{equation}
Evaluation of the jump across the
$R=R_{\cal K}$ surface gives
\begin{eqnarray}
\left({\partial \Psi^+ \over \partial R}
-{\partial \Psi^- \over \partial R}
\right)_{R_{\cal K}}&=& - \sum_{n=1,3,..}
a_n~{G_n(\zeta) \over R_{\cal K}^{n+1}}~
{2n+1 \over 1- \beta^{2n+1}}~,
\nonumber \\
&=&-{4\pi R_{\cal K} \sin \theta \over c}
{\cal K}_\phi(\zeta)~,
\end{eqnarray}
where $\beta \equiv R_*/R_{\cal K} <1$,
and ${\cal K}_\phi(\zeta)$ is the mentioned
surface current density.
    Thus we obtain
\begin{equation}
a_n ={4\pi R_{\cal K}^{n+2} \over c}~
{1-\beta^{2n+1} \over 2n( n+1)}
\int_{-1}^{1} d\zeta ~{{\cal K}_\phi(\zeta)G_n(\zeta)
\over (1-\zeta^2)^{1/2}}~.
\end{equation}
Note that $a_1$ is the contribution to
the star's dipole moment due to
the surface current ${\cal K}_\phi$,
and $a_3$ is related to the quadrupole
moment due to ${\cal K}_\phi$.
   Note also that $1-\beta^{2n+1} \approx
(2n+1)(\Delta R_a/R_*)/2$ which is
justified subsequently.

      From the considerations of \S 3,
we have
\begin{equation}
{d {\cal K}_\phi \over dt}
= \big(J_\phi v_R\big)_{R_a}~,
\end{equation}
where $J_\phi$ is given by equation (6).
Thus we find
\begin{equation}
{d a_n \over dt}=
{R_*^{n+1} v_{R*}\over 4n(n+1)}
(1-\beta^{2n+1})
\int_{-1}^{1} d\zeta~
{dP_n \over d\zeta}
{d H^2 \over d\Psi}\bigg|_{R_a}~,
\end{equation}
In particular,
\begin{equation}
{d a_1 \over dt}=
{1\over 4}{R_*^2 v_{R*}}(1-\beta^3)
\int_0^{\pi/2} d\theta~\sin\theta~
{d H^2 \over d\Psi}\bigg|_{R_a}~,
\end{equation}
which differs from equation (7)
by the factor $1-\beta^3$.
     Note that we have neglected the
gradual variation of the thickness $\Delta R_a$
of the accretion layer.  This is justified in \S 5.

    It is useful in subsequent work to introduce
\begin{equation}
A_1 \equiv \mu_i+a_1~,\quad A_n
\equiv a_n/R_*^{n-1}~,~~n=3,5,..
\end{equation}
   because
the $A_n's$ all have the same dimensions
as the dipole moment.

\subsection{Magnetic Energy}

     The magnetic energy of the system
is
\begin{eqnarray}
W&=&{1\over 8\pi}\int d^3 x~ ({\bf B}_i+{\bf B}^\prime)^2
=W_{i}+W^\prime,
\nonumber\\
W^\prime&=&{1\over 8\pi}\int d^3 x~ ({\bf B}^\prime)^2~.
\end{eqnarray}
Here, ${\bf B}_i$ is the initial dipole field and
$W_{i}$   the associated magnetic energy which is
independent of time.
Further, ${\bf B}^\prime(t)$ is the
field due to the currents on
the surface of the star which result from the accretion
and $W^\prime$ is the associated magnetic energy.
      The cross term involving
${\bf B\cdot B}^\prime$ in the first line
of the equation vanishes because in the volume in which ${\bf
B}^\prime$ is non-zero ($R \geq R_*$),  ${\bf B}_i$
can be written as the gradient of a potential.
      We can also write
$$
W^\prime ={1\over 2c}\int d^3x~
J_\phi^\prime A_\phi^\prime~.
$$
Using equation (16) and  equation
(13) then gives
\begin{equation}
W^\prime ={1 \over  R_*^3}{R_* \over \Delta R_a}
\sum_{n=1,3,..} {n(n+1) \over (2n+1)^2}~
(A_n-\mu_i\delta_{n0})^2~,
\end{equation}
where $\delta_{n0}$ is the Kronecker delta.
   We have made
the approximation
$1-\beta^{2n+1} \approx (2n+1)(\Delta R_a/R_*)/2$, which
is justified below where we estimate
$\Delta R_a$.

      Most of the magnetic energy $W^\prime$ is
in the accretion  layer on the star's surface.
    The reason for this is indicated by the schematic
drawing in Figure 3 of the fields ${\bf B}_i$
and ${\bf B}^\prime$.
    The magnetic
energy of the ${\bf B}^\prime$ field outside
of the star ($R\geq R_a$) is
\begin{eqnarray}
W^\prime({\rm ext}) &=&{1\over 4} \int_{-1}^1 d\zeta
\int_{R_a}^\infty R^2 dR ~
\left[(B_R^\prime)^2 + (B_\theta^\prime)^2\right]
\nonumber \\
&=& {1\over R_*^3} \sum_{n=1,3,..}{ n^2(n+1) \over 2 (2n+1)}
~(A_n-\mu_i\delta_{n0})^2,
\end{eqnarray}
where we have replaced $R_a$ by $R_*$ because $\Delta R_a
\ll R_*$.
     Retaining only the $n=1$ terms gives $W^\prime({\rm ext})/
W^\prime=3\Delta R_a/(2R_*) \ll 1$.

     The `buried' magnetic field is predominantly
parallel to the star's surface as indicated
by Figure 3.
    The magnetic energy $W^\prime$ of the
buried field, denoted $B^\prime_\parallel$,
  is approximately equal to
$4\pi R_*^2\Delta R_a (B_\parallel^\prime)^2/8\pi$.
    Retaining only the $n=1$ term in equation (23)
gives the estimate
\begin{equation}
B_\parallel^\prime \sim {2\over 3}
{R_* \over \Delta R_a} {|A_1-\mu_i| \over R_*^3}~.
\end{equation}
The buried field strength is  enhanced over
the initial dipole field ($\mu_i/R_*^3$)
by a factor $\approx R_*/\Delta R_a$ for $|A_1| \ll \mu_i$.

\subsection{Field at Large Distances from Star}

      For large $R$,
\begin{equation}
\Psi =  {A_1 G_1(\zeta)  \over
R} +{A_3 G_3(\zeta) R_*^2 \over
R^3} +...
\end{equation}
In the equatorial plane ($\zeta=0$),
\begin{equation}
\Psi(R,\pi/2)={A_1\over R} -
{3 A_3 R_*^2 \over 2R^3}+..,
\end{equation}
where
$A_1=\mu_i+a_1$
   is the
dipole moment at time $t$.
    We assume here and subsequently
that the decrease in
the dipole moment is limited
in the respect that
\begin{equation}
A_1 \gg {3|A_3| R_*^2 \over 2 r_A^2}~,
\end{equation}
and $A_1 \gg |A_n G_n(0)|(R_*/r_A)^{n-1}$
for $n=5,7,..$

Of course, the Alfv\'en radius
$r_A$ depends on the magnetic
field in the equatorial plane.
     This field is
\begin{equation}
B_\theta(R,\pi/2)=
-{A_1 \over R^3} +{9 A_3 R_*^2 \over 2R^5}+...
\end{equation}
Owing to inequality (28),
$|B_\theta(r_A,\zeta=0)| = |A_1|/r_A^3$ to a
good approximation.
     For simplicity, we consider
that $r_A$ depends only on
$\mu$ or $A_1$ and that the other quantities
in equation (2) such as $\dot{M}$
are constant.
     We then have
\begin{equation}
r_A =r_{Ai}\left({A_1\over \mu_i}\right)^{4/7}~,
\end{equation}
where $r_{Ai}$ is the Alfv\'en
radius at time $t=0$.
     Inequality (28) can then
be rewritten as
\begin{equation}
{A_1 \over \mu_i} \gg
\left|{A_3 \over \mu_i}\right|^{7/15}
\left({R_* \over r_{Ai}}\right)^{14/15}~.
\end{equation}
Owing to this inequality,
\begin{equation}
\Psi_{ff} ={\mu_i \over r_{Ai}}
\left({A_1 \over \mu_i}\right)^{3/7}~.
\end{equation}
We assume that $h_A/r_A=$ const $\ll 1$,
so that $\Delta \Psi_{ff}/\Psi_{ff}
=h_{Ai}/r_{Ai}$ = const $\ll1$.
    From \S 3 we have $H_m = A_1^2/r_A^4$.
Thus,
\begin{equation}
H_m^2 = {\mu_i^2 \over r_{Ai}^4}
\left({\mu_i \over A_1}\right)^{2/7}~.
\end{equation}

\subsection{Dimensionless Variables}

     It is natural to measure the
magnetic moment $A_1$ in units of
its initial value $\mu_i$,
\begin{equation}
\tilde{A}_1\equiv {A_1 \over
\mu_i}~,\quad{\rm and}
\quad\tilde{A}_n \equiv {A_n \over \mu_i}~,~n=3,5,..
\end{equation}
with the tildes indicating the dimensionless variables.
     We measure $\Psi$
in units of $\mu_i/R_*$, and
   $B_R(R_*,\zeta)$ in units
of $\mu_i/R_*^3$.
      In terms of the dimensionless variables,
   $\tilde{B}_R(R_*,\zeta)
= - d\tilde{\Psi}/d\zeta$.
   Further we have
\begin{equation}
\tilde{\Psi}_{ff}
={R_*\over r_{Ai}} ~\tilde{A}_1^{3/7}~,\quad {\rm and}\quad
\Delta \tilde{\Psi}_{ff}
={R_*\over r_{Ai}}{h_{Ai} \over r_{Ai}}~ \tilde{A}_1^{3/7},
\end{equation}
from equation (35).

     In subsequent work we drop the tilde's from the $A_n's$
and the $\Psi's$.

\begin{figure*}[t]
\epsscale{0.5}
\plotone{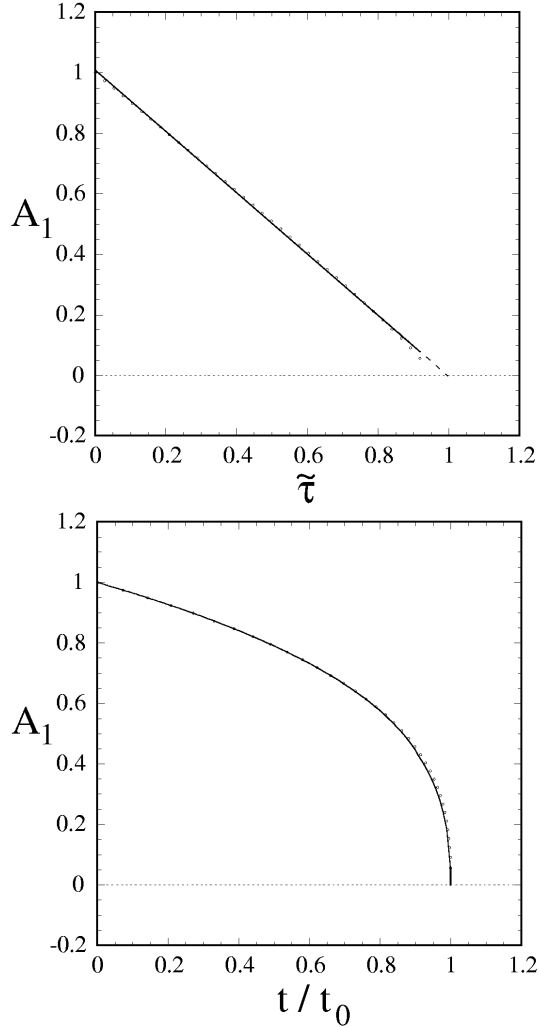}
\caption{Time dependence of the normalized  magnetic
moment $A_1/\mu_i$
  for the special case of only
the dipole mode.
     Here, $\mu_i$ is the initial
magnetic moment of the star, and $t_0$ is given
by equation (43).
  }
\end{figure*}

\subsection{Specific Model for $H(\Psi)$}

The qualitative dependence of $H(\Psi)$ is
discussed in \S 3.  A specific function
with this dependence is
\begin{equation}
\big[ H(\Psi)\big]^2 = H_m^2 \exp
\left(-{(\Psi-\Psi_{ff})^2 \over 2 \Delta \Psi_{ff}^2}\right)~.
\end{equation}
We can now rewrite equation (19) in dimensionless form as
\begin{equation}
{d {A}_n \over d\tau}=
-{2n+1 \over n(n+1)}~{{\cal I}_n \over {A}_1^{5/7}}~.
\end{equation}
    Here,
$$
{\cal I}_n \equiv \int_0^1 d\zeta ~{dP_n \over d \zeta}
X \exp\left(-{X^2 \over 2}\right),\quad
X(\zeta)\equiv
{\Psi(\zeta)-\Psi_{ff}\over \Delta \Psi_{ff}},
$$
where $\Psi(\zeta)\equiv \Psi(R_*,\zeta)$, with
$$
\Psi(\zeta)=\sum_{n=1,3,..} A_n G_n(\zeta)~.
$$
    The dimensionless time variable $\tau$ is given by
\begin{equation}
{d t \over d \tau} =
  {4}\left({ r_{Ai}\over R_*}\right)^3
{   R_* \over v_{R*}} ~{ \Delta  \Psi_{ff}\over \Psi_{ff} }
~{ R_* \over \Delta R_a(t)} ~.
\end{equation}
The initial conditions for equations (37) are $A_1(0)=1$ and
  $A_n(0)=0$ for $n=3,5,..$.

For simplicity, we consider $\Delta \Psi_{ff}/\Psi_{ff}$
to be a fixed number small compared with unity.
Consequently, the solution of equation (37)
depends on the dimensionless time $\tau$  and
on two dimensionless parameters, ${\cal E}$ and
$\epsilon$,
\begin{equation}
A_n=A_n(\tau,{\cal E},\epsilon)~,\quad
{\cal E} \equiv {R_* \over r_{Ai}}~, \quad
\epsilon \equiv {\Delta \Psi_{ff} \over \Psi_{ff}} =
{h_{Ai} \over r_{Ai}}~,
\end{equation}
where ${\cal E} \ll 1$ and $\epsilon \ll 1$.

\begin{figure*}[t]
\epsscale{0.5}
\plotone{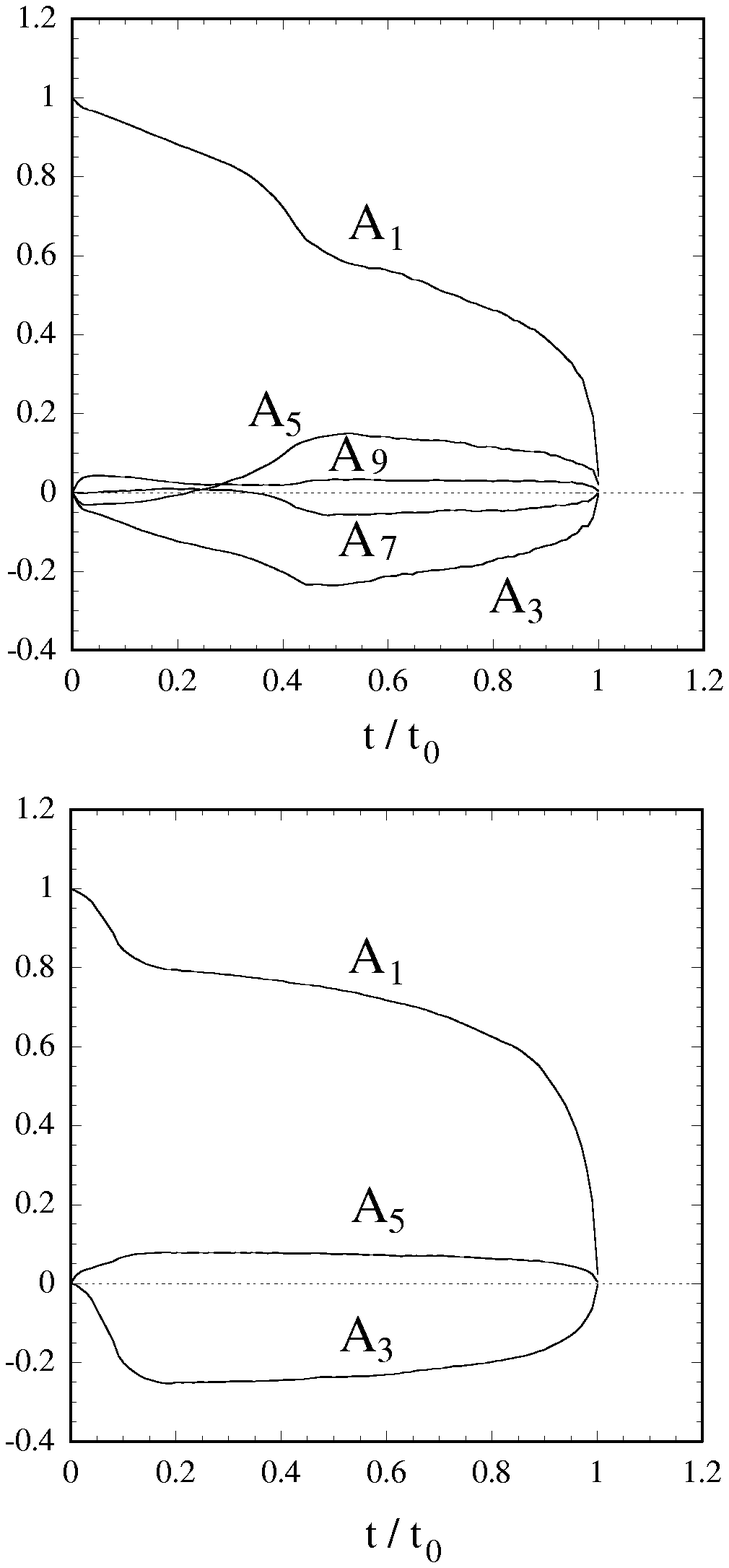}
\caption{Time dependence of the normalized  magnetic
multipole moments.  The top panel is
for a case with five modes for ${\cal E}=0.05$
and $\epsilon=0.1$, while the bottom panel is for
three modes for ${\cal  E}=0.025$ and $\epsilon=0.05$.
The time $t_0$ is given by equation (51).}
\end{figure*}

\subsection{Single Mode, $A_1 \neq 1$}

     For the special case of {\it only} one mode
$A_1 \neq 0$, equation (37) can be
approximated analytically.
   We find
\begin{equation}
{\cal I}_1 = - {A_1^{4/7} \over \epsilon ~{\cal E}}
\int_0^1 d\zeta ~(\zeta^2 - Z)\exp[-k(\zeta^2-Z)^2]~.
\end{equation}
Here,
$k\equiv A_1^{8/7}/(2\epsilon^2 {\cal E}^2)$ and
$Z\equiv 1-{\cal E}/A_1^{4/7}$, where we assume
that $Z>0$ or $A_1 >{\cal E}^{7/4}$.
   Because $\epsilon \ll 1$,
  $k \gg 1$ so that equation (40)
can be evaluated easily as
\begin{equation}
{\cal I}_1 \approx -{\sqrt{\pi/2} \over 2}
{\epsilon^2{\cal E}^2 \over
A_1^{8/7} Z^{3/2}}~.
\end{equation}
Thus we find
$$
{d A_1 \over d \tilde{\tau}} =-1~, \quad
{\rm or} \quad A_1=1-\tilde\tau~,
~ 0\geq\tilde \tau \leq 1~,
$$
\begin{equation}
{dt \over d\tilde{\tau}}=
{ 16\sqrt{2}\over 3\sqrt{\pi}}~
{ A_1^{13/7}\over \epsilon~{\cal E}^5 }~
{R_*\over v_{R*} }~
{ R_* \over \Delta R_a }~.
\end{equation}
We have used the approximation $Z=1$
which assumes $A_1 \gg {\cal E}^{7/4} \ll 1$.

     Equation (42) can be solved to give
$$
t=t_0[1-(1-\tilde\tau)^{20/7}]~.
$$
Here,
\begin{equation}
t_0 ={28 \sqrt{2} \over 15 \sqrt{\pi}}~
{R_* \over v_{R*}}~ {1 \over \epsilon~{\cal E}^5}~
{R_* \over <\!\!\Delta R_a \!>}~,
\end{equation}
is the {\it time-scale} for the field decrease.
This equation can also be written as
$$
t_0 \approx 1.5~
{R_* \over v_{R*}}~{r_{Ai} \over h_{Ai}}
\left({r_{Ai} \over R_*}\right)^5
{R_* \over <\!\!\Delta R_a \!>}~.
$$
Here, $<\!\!\Delta R_a\!\!>$
is the time-averaged thickness
of the accretion layer where it is assumed that
$\Delta R_a$ varies gradually with time.
   For a neutron star of mass $M=1.4M_\odot$
and radius $R_*=10^6$ cm, we find
\begin{equation}
t_0 \approx 0.78\times 10^5\left({0.001 \over \epsilon}\right)
\left({0.005 \over {\cal E}}\right)^5
\left({100 \over R_*/<\!\!\Delta R_a\!\!>}\right)~{\rm yr}~.
\end{equation}
The normalization of ${\cal E}$, or equivalently $r_{Ai}$,
follows from equation (2).  The normalization of $\epsilon$, or
equivalently $h_{Ai}/r_{Ai}$,  is based on the Shakura-Sunyaev
disk model (1973) which indicates that $\epsilon$ depends
quite weakly on the different parameters (e.g., $\epsilon
\propto \dot{M}^{1/5}$).  The normalization of
$ R_*/<\!\!\Delta R_a\!\!>$ is discussed in \S 5.
   The shortness of this time-scale is due to
the fact that only one mode is accounted for.

\subsection{Numerical Integrations}

   The results of \S 4.5 for a single mode
point up the importance  of using a new time
variable in place  of $\tau$ in equation (37).
    That is, we rewrite
this equation as
\begin{equation}
{d {A}_n \over d\tau^\prime}=
-{2n+1 \over n(n+1)}~A_1^{8/7}(\tau^\prime)~
{{\cal I}_n(\tau^\prime)
}~,
\end{equation}
for $n=1,3,..$,
where
\begin{equation}
{d \tau \over d \tau^\prime} = A_1^{13/7}(\tau^\prime)~.
\end{equation}
Combining this equation with equation (38) gives
\begin{equation}
{dt \over d\tau^\prime}=4~{\epsilon \over {\cal E}^3}~
{R_* \over v_{R*}}~
{R_* \over <\!\Delta R_a\!>}~ A_1^{13/7}~.
\end{equation}
   We let $\tau_0$ and $\tau_0^\prime$ denote
the values for which $A_1\leq 0.01$.
  We have
\begin{equation}
\tau_0 = \int_0^{\tau_0^\prime} d\tau^\prime
A_1^{13/7}(\tau^\prime)~.
\end{equation}
  The corresponding actual time when $A_1 \leq 0.01$ is
\begin{equation}
t_0=4~{\epsilon \over {\cal E}^3}~{R_* \over v_{R*}}~
{R_* \over <\!\Delta R_a\!>} ~\tau_0({\cal E},\epsilon)~.
\end{equation}
   For the case of a single mode ($A_1$),
$\tau^\prime$ is proportional to $\tilde{\tau}$ (neglecting the
weak time dependence of $\Delta R_a$).  For
this case $A_1$ decreases linearly with $\tilde{\tau}$
or $\tau^\prime$.

We find equations (45)  to be stiff.
   For this reason a
predictor-corrector code (Gear 1971) was developed and
used to solve the equations  for $1,~3,$ and $5$ modes.
The integrals ${\cal I}_n$ were evaluated using
a stretched $\zeta$ grid with $10^4-10^5$ points.
   Figure 4 shows the numerical solution for the case
of one mode ($A_1$).  This agrees closely with the analytic
solution of \S 4.5.

Figure 5 shows sample results for cases of three
and five modes.  The  time-scales
$t_0({\cal E},\epsilon)$ for three and five modes are not
appreciably different.
      Combining the results of many runs for
${\cal E}=0.01-0.1$ and $\epsilon = 0.02 -0.1$
we find
\begin{equation}
\tau_0 \approx {1.2 \times 10^{-4} \over
{\cal E}^{3.6}~ \epsilon^{2.7}}~.
\end{equation}
The integrations become increasingly long
as $\epsilon$ and ${\cal E}$ decrease.
In view of equation (49) we have
\begin{equation}
t_0 \approx 1.6\times 10^7\left({0.001 \over
\epsilon}\right)^{1.7}
\left({0.005 \over {\cal E}}\right)^{6.6}
\left({100 \over R_*/<\!\!\Delta R_a\!\!>}\right)~{\rm yr}~.
\end{equation}
This time-scale is considerably longer than that
for the case of the single mode (equation 44).
The time-scale  varies approximately as
$\mu_i^{3.8}/\dot{M}^{1.9}$.
   Thus for a neutron star with an initial
surface magnetic field of $2\times 10^{12}$ G
or $\mu_i=2\times 10^{30}$ Gcm$^3$, the
time-scale of equation (51) increases
to $2.2\times 10^8$ yr.

     Figure 6 shows the magnetic field lines
at a time during the evolution and the initial,
pure dipole field.
     Figure 7 shows the corresponding
surface current density
on the star's surface.

\section{Accretion Layer Thickness}

We now give a rough estimate the accretion layer
thickness $\Delta R_a$.
    We let $\Delta M$ denote the mass accreted
after  time $t$, which
is in a layer of thickness $\Delta R_a
\ll R_*$.   The mean density of the layer is
thus $\bar{\rho} \sim \Delta M/(4\pi R_*^2\Delta R_a)$.
The change in pressure across the layer in hydrostatic
equilibrium is
$ p \sim [\Delta M/(4\pi R_*^2)] g$, where
$g= GM/R_*^2 \approx 1.87\times 10^{14}$ cm/s$^2$.
   The equation of state of
the neutron star matter gives $p \approx k (\bar{\rho})^n$
with $k \approx 2.1\times 10^{11}$ cgs, and $n \approx 1.5$
(Baym, Bethe, \& Pethick 1971).
   Thus we obtain
\begin{eqnarray}
\Delta R_a &\sim& \left({k\over g}\right)^{1/n}
\left({\Delta M \over 4 \pi R_*^2}\right)^{(n-1)/n}
\nonumber \\
&\sim& 1.3\times10^4\left({\Delta M \over
10^{-2}M_\odot}\right)^{1/3}~{\rm cm}~.
\end{eqnarray}
For $\Delta R_a=10^4$ cm,
equation (25) gives an estimate of the
buried magnetic field of $\sim 6.7\times 10^{13}$ G.
Thus  the magnetic pressure in
the accretion layer is negligible compared
the matter pressure.

\subsection{Ohmic Diffusion of the Field}

    The influence of Ohmic diffusion on the
  `burial' of  magnetic field during
accretion to a neutron star has
been studied by Cumming, Zweibel,
and Bildsten (2001).
    They treat the competition between
the inward advection of the
magnetic field inside the
star and the tendency
of the field to  diffusion outward
due to the finite conductivity of
the plasma.
   For sufficiently
large accretion rates (such as that
considered here) the inward advection
is found to be larger than the
diffusion.

     Magnetic field buried during
intensive accretion will
  Ohmically  diffuse out of the star after
the accretion stops (BK74).
   The time scale for the buried field
to diffuse out of the star can be
estimated as
  \begin{equation}
t_{Ohm}\sim {(\Delta R_a)^2 \over \eta}~,\quad \quad
\eta ={c^2 \over 4\pi \sigma}~,
\end{equation}
where $\eta$ is the magnetic
diffusivity, and $\sigma$ is the conductivity.
Using  Cumming  et al.'s
\S 2.2 formula
for the conductivity and
our normalization values we find
\begin{equation}
t_{Ohm} \sim 2.7 \times 10^9
\left({\Delta R_a\over 10^4{\rm cm}}\right)
\left({\Delta M \over 10^{-2} M_\odot}\right)~{\rm yr}~.
\end{equation}
Thus it takes a rather long time for
the field to diffuse out of the star.

It is possible that the buried magnetic
field is subject to the interchange
or  buoyancy instabilities
considered by Cumming et al. (2001),
and by Kato, Fukue, and
Mineshige (1998),
but the detailed analysis is beyond the
scope of the present work.

\begin{figure*}[t]
\epsscale{0.8}
\plotone{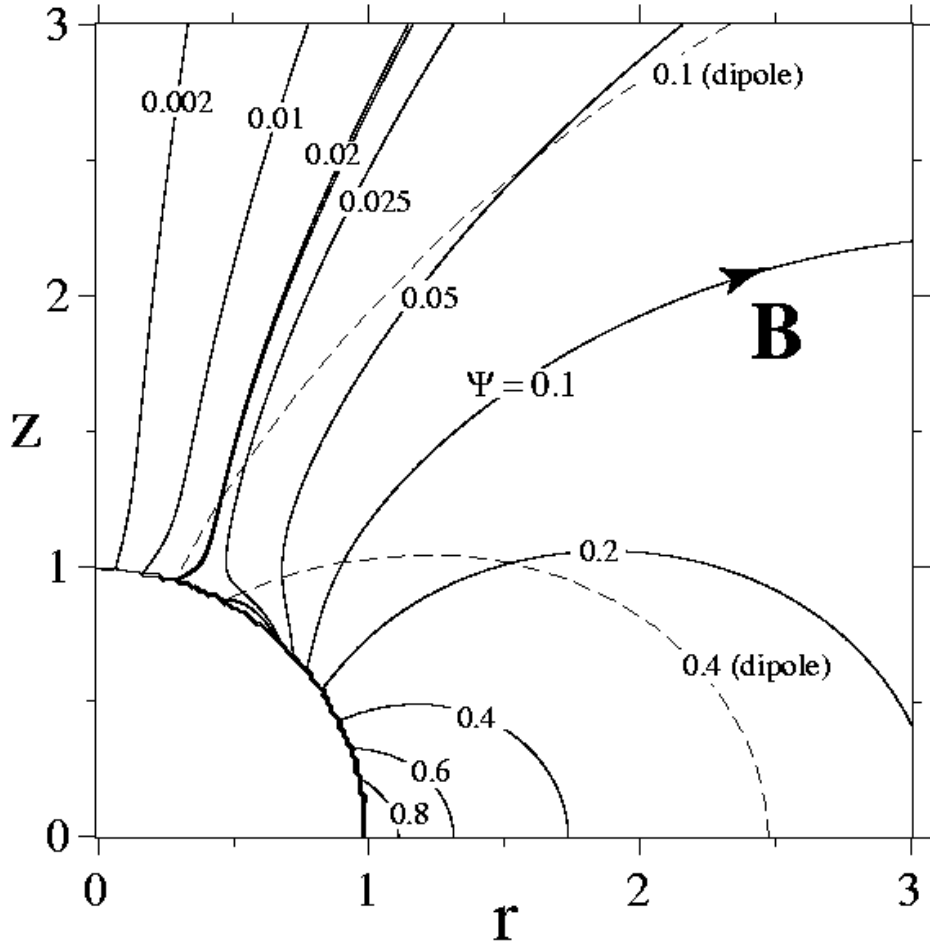}
\caption{Magnetic field lines of a screened
star (solid lines) with
three modes at $t/t_0=0.85$ when $A_1\approx 0.59$,
$A_3 \approx -0.19$,  and $A_5\approx 0.062$
for a case with ${\cal E}=0.025$ and $\epsilon=0.025$.
   The dashed lines are for the initial, pure dipole
field.
   The thicker line
$\Psi={\cal E}A_1^{3/7}\approx0.02$ is
the centroid of the funnel flow.
    It is clear from this figure that
the star's field is markedly changed
by the accretion.
}
\end{figure*}

\begin{figure*}[t]
\epsscale{1.}
\plotone{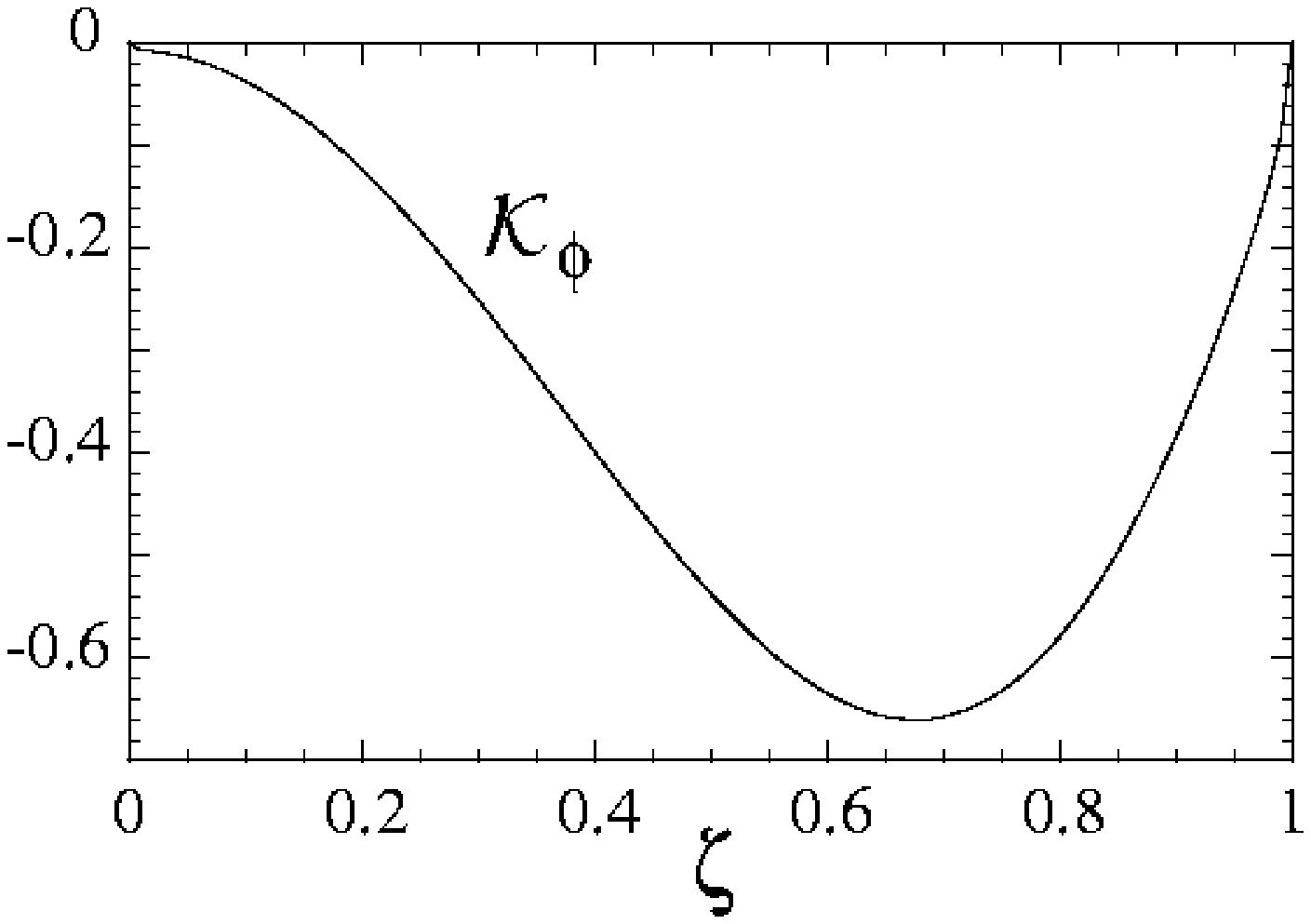}
\caption{The surface current density ${\cal K}_\phi$
is shown as a function of $\zeta =\cos(\theta)$,
with $\theta$ is the colatitude, for the same case as
Figure 6.  The vertical scale is arbitrary.}
\end{figure*}

\section{Conclusions}

     An analytical
model is developed for  the screening of
the external magnetic field of
a rotating, magnetized, axisymmetric neutron star
due to the  accretion of plasma
from a disk.
     The decrease of the
field occurs   due to the deposition
of current carrying plasma onto
the star's surface.
   This plasma  creates
an induced magnetic moment
with a sign opposite to that of the
original magnetic dipole.
    The physical mechanism is explained
in Figure 3.
    The equations (37) for the field evolution
are inherently nonlinear and this
leads to the generation of higher
order modes when starting from
the lowest order mode.

   The main conclusions from this
work are:
   (1) The  field decreases
independent of whether the star
spins-up or spins-down (that is,
for $r_{cr} > r_A$ or $r_{cr} < r_A$);
   (2) The time-scale for an appreciable
decrease (by a factor $>100$)
of the field is $t_0  \sim 1.6 \times 10^7$ yr
for $\dot{M}=10^{-9} M_\odot$/yr and
an initial stellar magnetic moment
$\mu_i=10^{30}$ Gcm$^3$
and it scales approximately
  as $\mu_i^{3.8}/\dot{M}^{1.9}$;
   (3) The decrease of the magnetic field
does not have a simple relation to
the accreted mass;
(4) At late times
the magnitude of the
buried magnetic field
is much larger than the initial
field on the star's surface;  and (5)
   Once the accretion stops the field
leaks out on an Ohmic diffusion time scale
which is estimated to be $\gtrsim 10^9$ yr.

    The present model has evident  limitations
which require further study:
Accreting neutron stars are not
expected to have their magnetic
and rotational axes aligned.
    The case of small misalignment angles
may be amenable to analytic treatment.
   The thickness of the accretion layer
will in general be a function of
position.

{We thank Dr. Dong Lai
for a number of valuable discussions.
    We also thank the referee for valuable
recommendations.
This work was supported in part by NASA grant
  NAG 5-13220 and NSF grant AST-0307817.}

\end{document}